\documentclass{ws-ijmpa}
\usepackage[super,compress]{cite}
\usepackage{graphicx}
\begin{document}
\markboth{F.Taghavi-Shahri,S.Atashbar Tehrani,M.Zarei}{Fragmentation Functions of neutral  mesons $\pi^0$ and $k ^0$ with Laplace transform approach}

%
\catchline{}{}{}{}{}
%

\title{Fragmentation Functions of neutral  mesons $\pi^0$ and $k ^0$ with Laplace transform approach
}

\author{F.Taghavi-Shahri\footnote{Corresponding author: taghavishahri@um.ac.ir.}
}

\address{Department of Physics, Ferdowsi University of Mashhad\\
Mashhad, Iran, P.O. Box 1436\\taghavishahri@um.ac.ir}

\author{S.Atashbar Tehrani}

\address{Independent researcher\\
P.O. Box 1149-8834413 Tehran, Iran\\atashbar@ipm.ir}

\author{M.Zarei}

\address{Department of Physics, Ferdowsi University of Mashhad\\
	Mashhad, Iran, P.O. Box 1436\\ m\_zarei\_128@yahoo.com}

\maketitle

\begin{history}
\received{Day Month Year}
\revised{Day Month Year}
\end{history}

\begin{abstract}
With an  analytical solutions of DGLAP evolution equations based on the Laplace transform method , we find  the fragmentation functions (FFs) of neutral mesons, $\pi^0$ and $k ^0$ at NLO approximation.  We also calculated the total fragmentation
functions of these mesons and compared them with experimental
data and those from global fits. The results show a good agreements   between our solutions and other models and also are compatible  with experimental data.

\end{abstract}

\ccode{PACS numbers:12.38.Bx, 13.60.Hb, 13.85.Hd, 13.66.Bc}


\section{Introduction}

Fragmentation process is the QCD process in which partons hadronize to colorless  hadrons. In this transition, the parton
fragmentation function, $D_i^h (z,Q^2)$,   represents the probability for a parton $i$ to fragments into
a particular hadron $h$ carrying a certain fraction of the parton energy or momentum.
Therefore, these fragmentation functions (FFs) are  essential inputs to study the hadron production in any
processes like $p\bar{p}$, $ep$, $\gamma p$ and $\gamma \gamma$ scattering.\\
Fragmentation Functions  evolved with DGLAP evolution equations from a starting distribution at a defined energy scale \cite{HALZEN},\cite{split}.\\
Recently we have used Laplace transformation and provided an
analytical solution to DGLAP equations  to calculate the proton, pion and kaon fragmentation functions \cite{taghavi0}.
This method had  provided  analytical solutions to Polarized Parton Distribution
Functions (PPDFs) too\cite{taghavi1,taghavi2}.\\
In the present paper we will used the results of this new method introduced by Block et
al.\cite{Block1,Block2,Block3,Block4,Block5,Block6} to calculate
the neutral  mesons,$\pi^0$ and $k^0$ fragmentation functions. \\
Therefore, our main task here is using our solutions for charged pion and kaon fragmentation functions \cite{taghavi0}, for calculating
the neutral  mesons, $\pi^0$ and $k ^0$ fragmentation functions. These solutions enable us to use the neutral mesons data in a global fit, in addition to  all experimental data for total fragmentation functions of charged mesons, to determination of FFs.\\
The paper is organized as follows.  In Section 2 we review  the Laplace transform method of non- singlet, singlet and gluon DGLAP evolution equations for extracting the fragmentation functions. These solutions led us to $\pi^+$ and $k^+$  FFs. Then, in  Section 3  we utilize the charge conjugation symmetry to calculate   the fragmentation functions  of $\pi^-$ and $k^-$. This led us to natural mesons fragmentation functions.
Finally in section 4 we   calculated the  fragmentation  functions of neutral  mesons, $\pi^0$ and $k ^0$ and also compared them with available experimental  data \cite{tasso,aleph,topaz,opal} and those from global fits \cite{hira1,hira2,kkp1,kkp2,DSS}.

\begin{figure}
	\begin{center}
	\includegraphics[clip,width=0.75\textwidth]{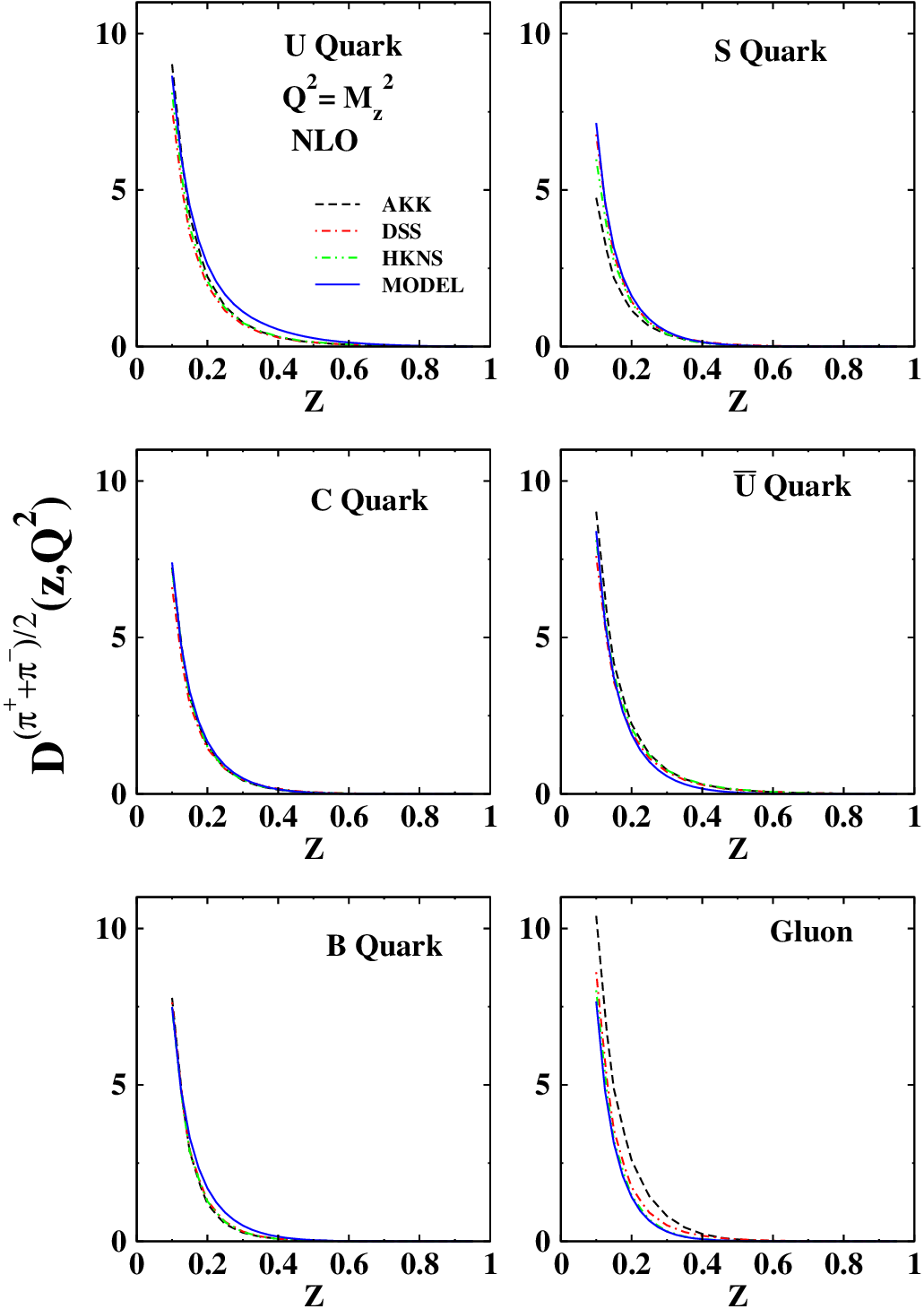}
		\caption{Fragmentation functions of natural  pion and comparison  with  AKK, DSS and HKNS global fits.                              }\label{fig:figQ0}
	\end{center}
\end{figure}

\begin{figure}
	\begin{center}
	\includegraphics[clip,width=0.75\textwidth]{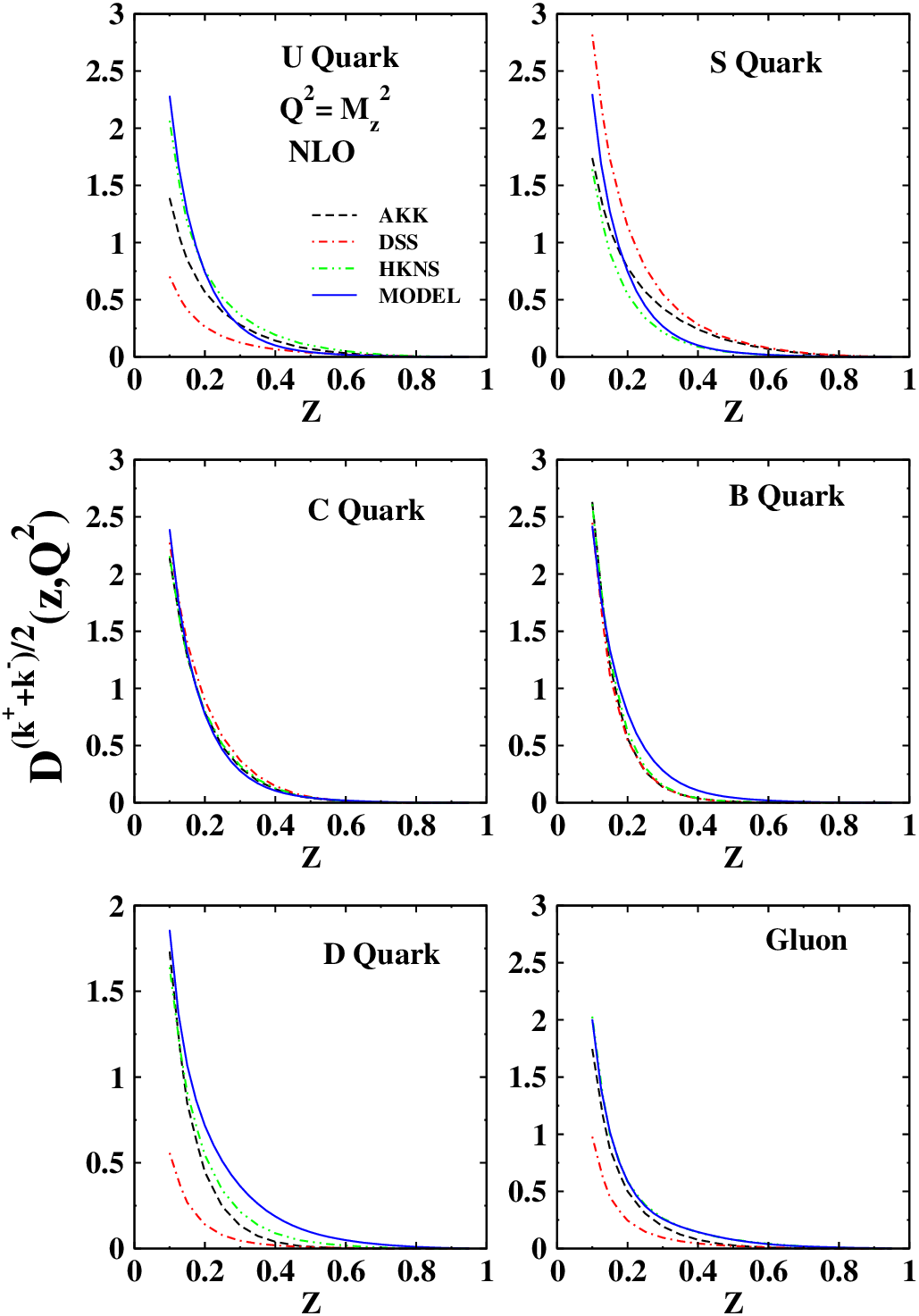}
		\caption{ Fragmentation functions of natural   kaon and comparison  with AKK, DSS and HKNS global fits. }\label{fig:figQ0}
	\end{center}
\end{figure}

\section{Fragmentation Functions via decoupling of the DGLAP evolution equations by Laplace
	transform method }

\subsection{Non- singlet case:}

The fragmentation of valence quarks  into
hadrons defined the  non- singlet fragmentation functions. The evolution of non- singlet fragmentation function are given by  DGLAP evolution equations at NLO approximation as:

\begin{eqnarray}
	{4\pi\over \alpha_s(Q^2)}{\partial D_{ns}\over \partial \ln
		(Q^2)}(z,Q^2)&=&D_{ns}\otimes\left[
	P_{qq}^{LO,ns}+{\alpha_s(\tau)\over
		4\pi}P_{qq}^{NLO,ns}\right](z,Q^2).\label{nonsingletinQsq_1}
\end{eqnarray}
where
\begin{equation}
	D_{ns}^h (z,Q^2)= D_q^h
	(z,Q^2) -  D_{\bar q}^h (z,Q^2)
\end{equation}
The  $\otimes$ symbol in  Eq. (1) refers to the
convolution integral. In the new method introduced by  Block et
al.\cite{Block1,Block2,Block3,Block4,Block5,Block6} ,The  DGLAP evolution equations can be solved by Laplace transform approach.
To summarized, by introducing two variable $\nu\equiv
ln(\frac{1}{z})$ and $\tau \equiv \frac{1}{4\pi
}\int_{Q_{0}^{2}}^{Q^{2}}\alpha _{s}(Q^{\prime 2})d\ln Q^{\prime
2}$, and two Laplace transforms from $\nu$ space to $s$
space and from $\tau$ space to $U$ space,  the DGLAP evolution equations can be solved
iteratively  by  a set of convolution integrals which are related
to initial input Fragmentation Functions  at  scale of $Q_0^2$.
Two inverse Laplace transforms will take us back to the
usual space ($z$, $Q^2$)\cite{taghavi0}. We defined  $zD_{ns}(z,Q^2)=F_{ns}(z,Q^2)$, and  find  the solution of non- singlet DGLAP evolution
equation, Eq. (1) in $s$ space as \cite{taghavi0} :

\begin{equation}
	f_{ns}(s,\tau)=e^{\tau
		\Phi_{ns}(s)}f_{ns0}(s),\label{solutionandPhinsofs}
\end{equation}
where
\begin{equation}
	\Phi_{ns}(s)\equiv \Phi_{ns}^{LO}(s)+{\tau_2\over
		\tau}\Phi_{ns}^{NLO}(s),
\end{equation}

$\Phi_{ns}^{LO}(s)$ and $\Phi_{ns}^{NLO}(s)$ are the Laplace transform of non- singlet splitting functions and are given in
Appendix. A of \cite{taghavi0}. The  $\tau_2$ parameter  is defined as
\begin{eqnarray}
	\tau_2&\equiv& {1\over 4\pi}\int_0^\tau\alpha_s(\tau')\,d\tau'
	={1\over (4\pi)^2}\int_{Q^2_0}^{Q^2}\alpha_s^2(Q'^2)\, d \ln
	Q'^2,\label{tau2ofQsq}
\end{eqnarray}

$f_{ns0}(s)$ in Eq. (3) is the Laplace transform of initial valence quark
fragmentation functions at $Q_0=4.5 GeV$. They are selected from HKNS code \cite{hira1} to be sure about our analytical solutions of DGLAP evolution equation. Finally, with an inverse Laplace transform  of Eq. (3)\cite{Block6}, one can derive
the valence quark fragmentation functions in $(z,Q^2)$ space.

\subsection{The singlet and gluon case:}

At NLO approximation, the singlet and gluon fragmentation functions are given by these two coupled DGLAP evolution equations:
\begin{equation}
	\frac{4\pi}{\alpha_s(Q^2)}\frac{\partial
		D_s}{\partial\ln{Q^2}}(z,Q^2)= D_s\otimes \left(
	P_{qq}^0+\frac{\alpha_s(Q^2)}{4\pi}P_{qq}^1\right)(z,Q^2)+D_g
	\otimes \left(
	P_{gq}^0+\frac{\alpha_s(Q^2)}{4\pi}P_{gq}^1\right)(z,Q^2),\label{DS}
\end{equation}

\begin{equation}
	\frac{4\pi}{\alpha_s(Q^2)}\frac{\partial D_g }{\partial
		\ln{Q^2}}(z,Q^2)= D_s\otimes \left(
	P_{qg}^0+\frac{\alpha_s(Q^2)}{4\pi}P_{qg}^1\right)(z,Q^2)+D_g
	\otimes \left(
	P_{gg}^0+\frac{\alpha_s(Q^2)}{4\pi}P_{qg}^1\right)(z,Q^2).\label{DG}
\end{equation}
where the singlet fragmentation function, $D_s^h (z,Q^2)$ , is defined as
\begin{equation}
	D_s^h (z,Q^2)=\sum_{q={u,d,s,c,b}} [ D_q^h (z,Q^2) +  D_{\bar q}^h
	(z,Q^2) ]
\end{equation}
By definition of  $z D_{s}(z,Q^2) \equiv F_{s}(z,Q^2)$ and $z
D_{g}(z,Q^2) \equiv G(z,Q^2)$,  the solutions of these coupled DGLAP evolution equations in Laplace $(s,U)$  space can be calculated as \cite{taghavi0}:

\begin{eqnarray}
	\left[ U -\Phi_f(s)\right]{\cal F}(s,U)&-& \Theta_g(s){\cal G}(s,U)= f_0(s)\nonumber\\
	& +& a_1\left[\Phi_f^{NLO}(s){\cal F}(s,U+b_1)+\Theta_g^{NLO}(s)
	{\cal G}(s,U+b_1)\right] ,\label{Feqn}
\end{eqnarray}

\begin{eqnarray}
	-\Theta_f(s){\cal F}(s,U)&+&\left[ U -\Phi_g(s)\right]{\cal G}(s,U)= g_0(s)\nonumber\\
	&+& a_1\left[\Theta_f^{NLO}(s){\cal
		F}(s,U+b_1)+\Phi_g^{NLO}(s){\cal G}(s,U+b_1)\right] ,\label{Geqn}
\end{eqnarray}
here the ${\cal F}(s,U)$ and ${\cal G}(s,U)$ are the Laplace transformed of singlet and gluon fragmentation functions in $(s,U)$ space. Initial input fragmentation functions are denoted by  $f_0(s)$ and  $g_0(s)$.  As we mentioned before, these  initial inputs are selected  from HKNS code
\cite{hira1} at initial scale of $Q_0=4.5 GeV$. The parameters of $a1 = 0.025$ and $ b1 = 10.7$ are the best fit parameters to $a(\tau)={\alpha_s(\tau)\over 4\pi} \approx a_0+ a_1e^{-b_1\tau}$ at NLO approximation \cite{Block1}.
The functions  $\Phi_{f,g}$ and $\Theta_{f,g}$  specified the laplace transformation of splitting functions and can be found in \cite{taghavi0} and also given in Appendix A:
\begin{eqnarray}
	\Phi_f(s)\equiv \Phi_f^{LO}(s)+a_0\Phi_f^{NLO}(s),\qquad
	\Phi_g(s)\equiv \Phi_g^{LO}(s)+a_0\Phi_g^{NLO}(s),\label{Phis}
\end{eqnarray}
\begin{eqnarray}
	\Theta_f(s)\equiv \Theta_f^{LO}(s)+a_0\Theta_f^{NLO}(s),\qquad
	\Theta_g(s)\equiv
	\Theta_g^{LO}(s)+a_0\Theta_g^{NLO}(s),\label{Thetas}
\end{eqnarray}
With the initial input functions for $f_0(s)$ and $g_0(s)$, their evolved solutions in
the Laplace s space are given by \cite{Block5}
\begin{eqnarray}
	f(s,\tau)=k_{ff}(s,\tau)f_0(s)+ k_{fg}(s,\tau)g_0(s)\nonumber\\
	g(s,\tau)=k_{gg}(s,\tau)g_0(s)+ k_{gf}(s,\tau)f_0(s)
\end{eqnarray}
where the $ k$'s  in Eq. (13) are  given in Appendix. B of Ref. \cite{taghavi0} for the
first iteration.  Finally, the singlet and gluon
fragmentation functions in $(z,Q^2)$ space can be calculated  with known inverse laplace transform \cite{Block6}.
Our results in Ref. \cite{taghavi0} show a nice agreement between these analytical solution  and  other global fits results for charged mesons $\pi^+$ and $k ^+$.

\section{ Natural pions and kaons fragmentation functions and the role of charge conjugation symmetry }
The fragmentation function of total sea quarks  is defined as follows
\begin{eqnarray}
	D_{s}(z,Q^2)- D_{ns}(z,Q^2)= D_{\bar q}(z,Q^2)
\end{eqnarray}
Where $ D_{\bar q}(z,Q^2)$ is
\begin{eqnarray}
	D_{\bar q}(z,Q^2)=2 D_{\bar u}(z,Q^2)+2 D_{\bar d}(z,Q^2)+2 D_{s}(z,Q^2)+2 D_{c}(z,Q^2)+2 D_{b}(z,Q^2),\nonumber\\
\end{eqnarray}
Because  the  heavier sea quarks can produce hadrons with higher
probability, we simply parameterized the fraction of  different kind of sea
quarks fragmentation functions  as their mass ratio and then we have\cite{taghavi0}:
\begin{eqnarray}
	D_{quark} (z,Q^2)=\frac{D_{\bar q}(z,Q^2)}{B^A},
\end{eqnarray}
The parameters of $A$ and $B$ are summarized in Table 1 in \cite{taghavi0}.
We also used these flavor  symmetries between different kinds of
fragmentation functions in $\pi^+$, $K^+$ \cite{hira1}:
\begin{eqnarray}
	D_{\bar u}^{\pi^+} (z,Q^2)  = D_{d}^{\pi^+} (z,Q^2)
	\ne D_{s}^{\pi^+} (z,Q^2)\nonumber\\
	D_{u}^{\pi^+} (z,Q^2) = D_{\bar d}^{\pi^+} (z,Q^2)\nonumber\\
	D_{s}^{\pi^+} (z,Q^2) = D_{\bar s}^{\pi^+} (z,Q^2)\nonumber\\
	D_{c}^{\pi^+} (z,Q^2) = D_{\bar c}^{\pi^+} (z,Q^2)\nonumber\\
	D_{b}^{\pi^+} (z,Q^2) = D_{\bar b}^{\pi^+} (z,Q^2)
\end{eqnarray}
\begin{eqnarray}
	D_{\bar u}^{K^+} (z,Q^2) \ne D_{d}^{K^+} (z,Q^2)
	\ne D_{s}^{K^+} (z,Q^2)\nonumber\\
	D_{d}^{K^+} (z,Q^2) = D_{\bar d}^{K^+} (z,Q^2)\nonumber\\
	D_{c}^{K^+} (z,Q^2) = D_{\bar c}^{K^+} (z,Q^2)\nonumber\\
	D_{b}^{K^+} (z,Q^2) = D_{\bar b}^{K^+} (z,Q^2)
\end{eqnarray}
To calculate the natural mesons fragmentation functions, we first used the charge conjugation symmetries related the $\pi^+$($K^+$) fragmentation functions  to those of $\pi^-$($K^-$) to derive the $\pi^-$($K^-$) fragmentation functions:
\begin{eqnarray}
	D_{\bar q}^{\pi^+ (K^+)} (z,Q^2)  = D_{q}^{\pi^-(K^-)}(z,Q^2)\nonumber\\
	D_{g}^{\pi^+(K^+)} (z,Q^2)  = D_{g}^{\pi^-(K^-)}(z,Q^2)
\end{eqnarray}
Finally the neutral mesons fragmentation functions can be obtained by:
\begin{eqnarray}
	D_{i}^{\pi^0} (z,Q^2) = \frac{1}{2} [D_{i}^{\pi^+} (z,Q^2) + D_{i}^{\pi^-} (z,Q^2)]\nonumber\\
	D_{i}^{K^0} (z,Q^2) = \frac{1}{2} [D_{i}^{K^+} (z,Q^2) + D_{i}^{K^-} (z,Q^2)]
\end{eqnarray}

Figures (1) and (2) show the results of fragmentation functions of neutral pions and kaons at $Q^2=M_z^2 GeV^2$. We also compared our results with those of AKK, DSS and HKNS codes\cite{hira1,kkp2,DSS}  to be sure about our analytical solutions.

\section{ Total Fragmentation Functions}

According to the factorization theorem \cite{collins}, the total fragmentation function can be
expressed in terms of the partonic hard scattering cross
sections and the non-perturbative fragmentation functions as:
\begin{eqnarray}
	F^H(z,Q^2)=\frac{1}{\sigma_{tot}}\frac{d\sigma^h}{dz}(e^- e^+ \rightarrow H X)
	(z,Q^2)=\sum C_i(z,Q^2) \otimes D_i^H(z,Q^2)
\end{eqnarray}
where, $\sigma_{tot}$ is the total hadronic cross section \cite{kkk}. The  $C_i(z,Q^2)$ is the Wilson  coefficient function related to the partonic cross section $e^- e^+ \rightarrow q \bar q$ and calculated in perturbative QCD as\cite{kkp1,wilson}:

\begin{eqnarray}
	C_q^1 (z)=C_F\left[(1+z^2)
	\left(\frac{\ln(1-z)}{1-z}\right)_{+}-\frac{3}{2}\frac{1}{(1-z)_{+}}
	+
	2\frac{1+z^2}{1-z}\ln(z)+\frac{3}{2}(1-z)+\left(\frac{3}{2}\pi^2-\frac{9}{2}\right)
	\delta(1-z) \right],\nonumber\\
\end{eqnarray}

\begin{eqnarray}
	C_g^1 (z)&=&2 C_F \left[\frac{1+(1-z)^2}{z}\ln(z^2 (1-z))
	-2\frac{1-z}{z}\right],
\end{eqnarray}

\begin{eqnarray}
	C_q^L (z)&=&C_F,
\end{eqnarray}

\begin{eqnarray}
	C_g^L (z)&=&4 C_F\frac{(1-z)}{z}.
\end{eqnarray}
where $ C_F=\frac {4}{3}$. The total fragmentation functions of natural pion,$\pi ^0$ and kaon, $K^0$  are shown in Fig. (3) and Fig. (4) . We compared our result with
those from HKNS global fit and also with data from TASSO, ALEPH,TOPAZ and  OPAL  experiments
\cite{tasso,aleph,topaz,opal}. The agreement between experimental data and our model is quite
reasonable. This means that our analytical solutions are correct.

\begin{figure}
	\begin{center}
	\includegraphics[clip,width=0.55\textwidth]{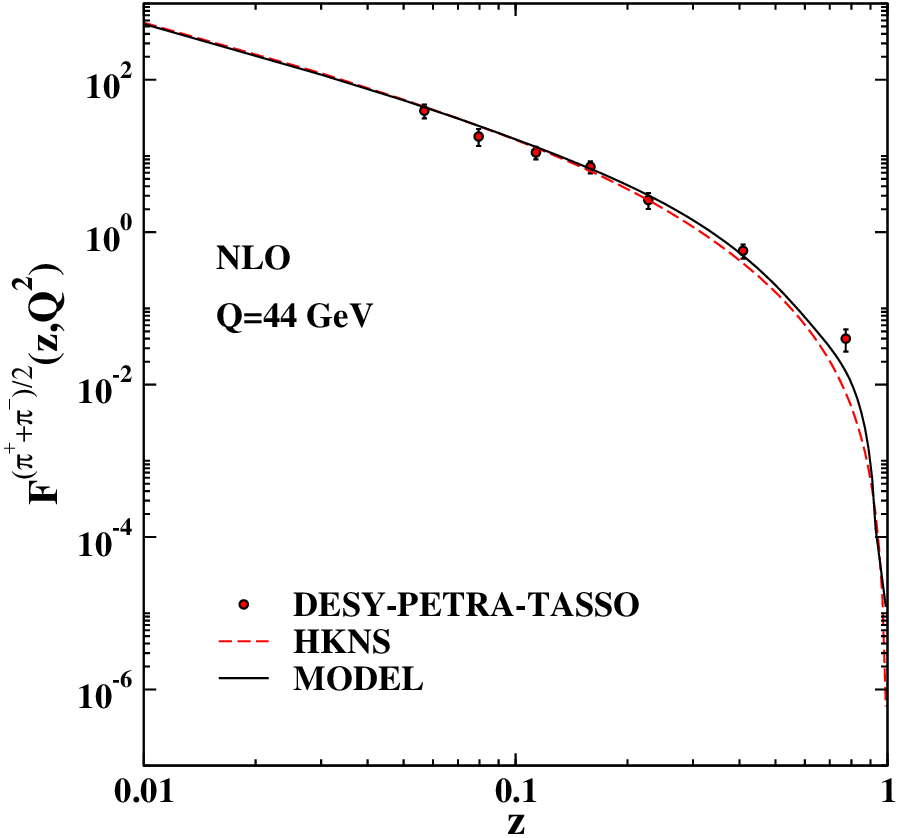}
	\includegraphics[clip,width=0.55\textwidth]{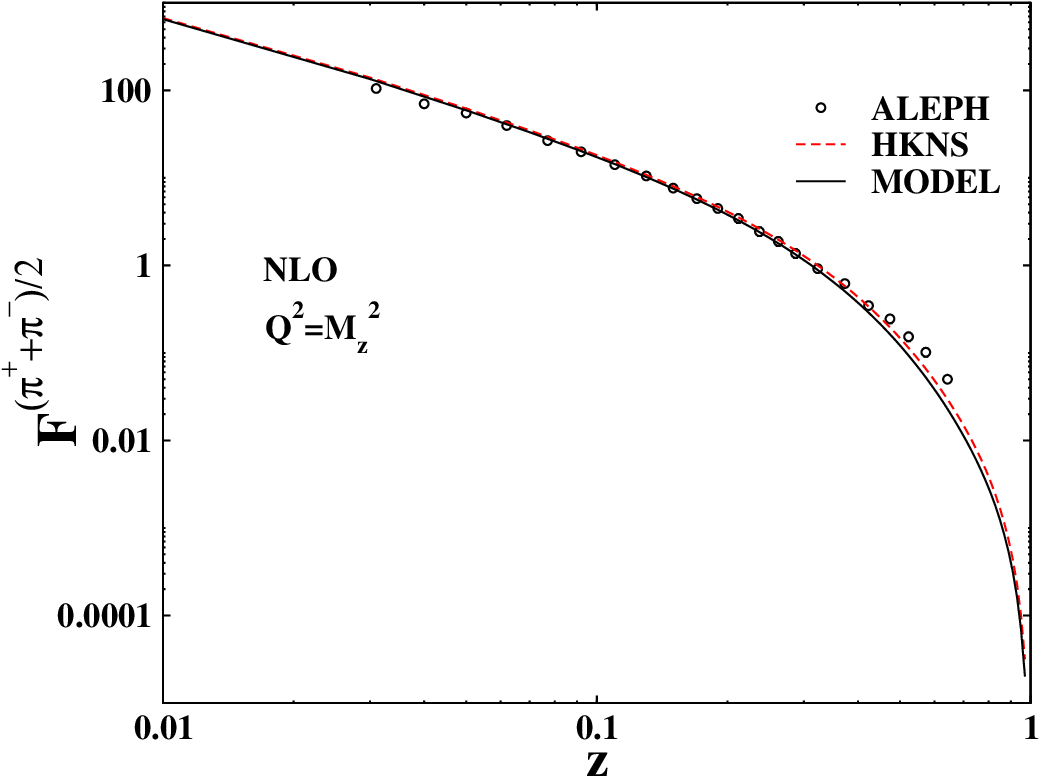}
		\caption{Total fragmentation functions of natural pion
			and comparison with experimental data from TASSO and ALEPH Collaborations\cite{tasso,aleph} at $Q=44 GeV$ and
			$Q^2=M_z^2$. We also compared our results with
			HKNS global fit. }\label{fig:figQ0}
	\end{center}
\end{figure}

\begin{figure}
	\begin{center}
	\includegraphics[clip,width=0.55\textwidth]{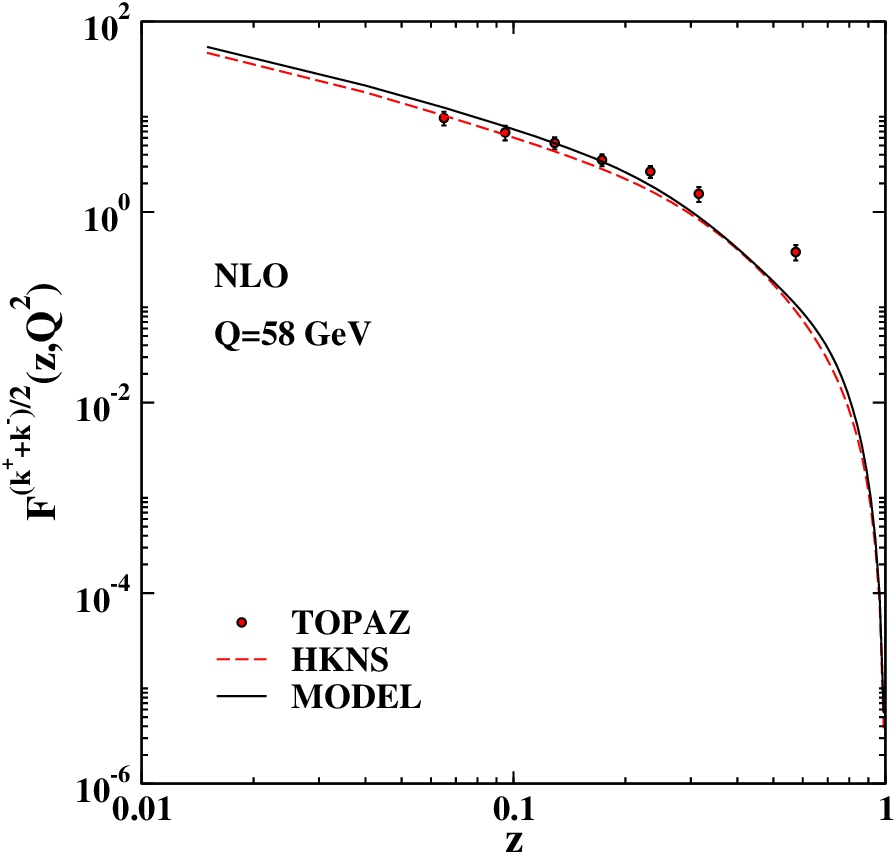}
	\includegraphics[clip,width=0.55\textwidth]{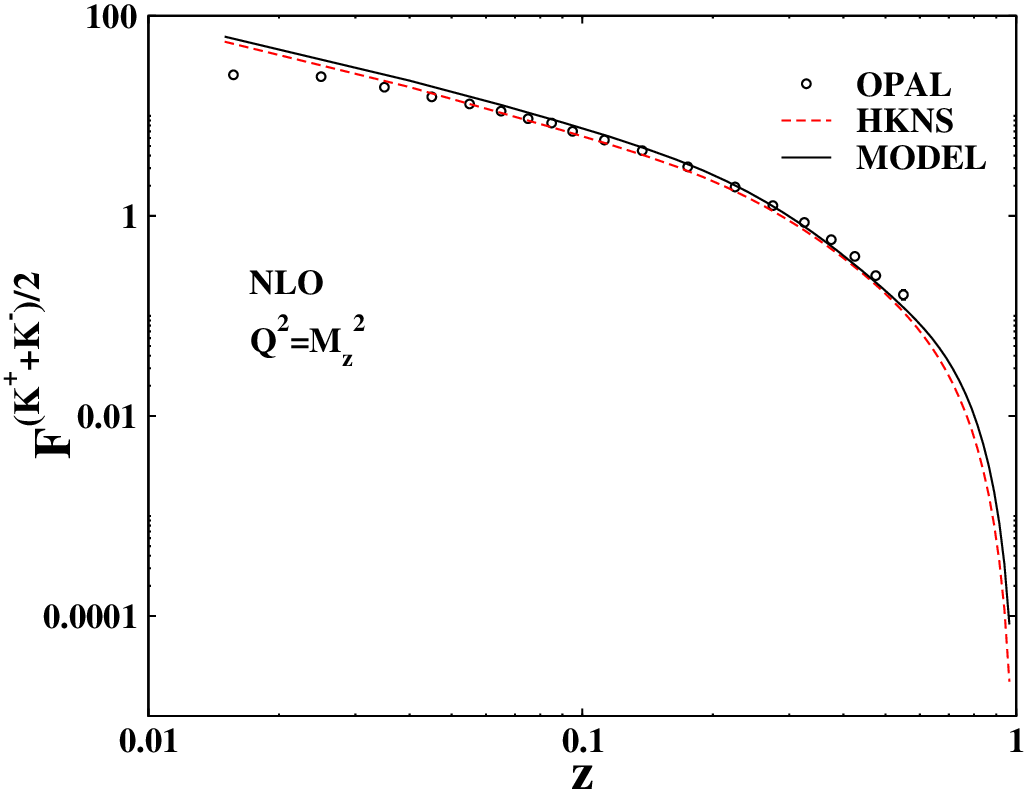}
		\caption{Total fragmentation functions of natural  kaon and proton
			and comparison with experimental data from TOPAZ and  OPAL Collaborations \cite{topaz,opal} at $Q=44 GeV$ and
			$Q^2=M_z^2$. We also compared our results with
			HKNS global fit.    }\label{fig:figQ0}
	\end{center}
\end{figure}

\section{Conclusions and Remarks}
Using the analytical solutions to DGLAP evolution equation,based on the Laplace transforms, we find the fragmentation functions of the neutral pions and kaons. Finding these solutions enable us to use the natural mesons experimental data for total fragmentation functions in a global fit to determination of fragmentation functions. This technique has the  facility that the
analytical solution of the fragmentation functions are
obtained more strictly by using the related kernels and the calculations controlled  in a better way.
We have  used the HKNS code for initial input fragmentation
functions to be sure about our analytical solutions.\\
Our results for natural pions and kaons are compared with those from global fits and also with
experimental data  and  there is a reasonable agreements between them.

\section*{ Acknowledgment}

This work is supported by Ferdowsi University of Mashhad under
grant 2/39420(1394/11/04).

\appendix

\section*{Apendix A}
We present here the results for the Laplace
transforms of   splitting  functions, denoted by $\Phi^{LO,NLO}$ and $\Theta^{LO,NLO}$ at the
NLO approximation.

\begin{eqnarray*}
	\Phi^{LO}_f(s)&=&4-\frac{8}{3}\left(\frac{1}{s+1}+\frac{1}{s+2}+2(\psi(s+1)+\gamma_E)\right),\nonumber\\
	\Theta^{LO}_g(s)&=&\frac{16}{3} n_f \left(\frac{2}{s}-\frac{2}{s+2}+\d frac{2}{s+3}\right),\nonumber\\
	\Theta^{LO}_f(s)&=&\frac{1}{s+1}-\frac{2}{s+2} + \frac{2}{s+3},\nonumber\\
	\Phi^{LO}_g(s)&=&\frac{33-2 n_f}{3}+12 \left(\frac{1}{s}-\frac{2}{s+1}+\frac{1}{s+2}-\frac{1}{s+3}-\psi(s+1)-\gamma_E\right),\nonumber
\end{eqnarray*}

\begin{eqnarray*}
	&&\Phi _{nsqq}^{NLO} =\\
	&&C_{F}T_{f}\left( -\frac{2}{3(s+1)^{2}}-\frac{2}{9(s+1)}%
	-\frac{2}{3(s+2)^{2}}+\frac{22}{9(s+2)}+\frac{4}{3}\psi ^{\prime
	}(s+1)\right) + \\
	&&C_{F}^{2}\left( \frac{5}{(s+1)^{3}}+\frac{5}{(s+1)^{2}}-\frac{5}{s+1}+%
	\frac{5}{(s+2)^{3}}+\frac{3}{(s+2)^{2}}+\frac{5}{s+2}\right.  \\
	&&\left. -\frac{2}{(s+1)^{2}}\left( \gamma _{E}+\frac{1}{s+1}\psi
	(s+1)-(s+1)\psi ^{\prime }(s+2)\right) \right.  \\
	&&\left. -\frac{2}{(s+2)^{2}}\left( \gamma _{E}+\frac{1}{s+2}\psi
	(s+2)-(s+2)\psi ^{\prime }(s+3)\right) \right.  \\
	&&\left. +4\left( \left( \psi (s+1)+\gamma _{E}\right) \psi ^{\prime }(s+1)-%
	\frac{1}{2}\psi ^{\prime \prime }(s+1)\right) -3\psi ^{\prime
	}(s+1)+4\psi
	^{\prime \prime }(s+1)\right)  \\
	&&+C_{A}C_{F}\left( -\frac{1}{(s+1)^{3}}+\frac{5}{6(s+1)^{2}}+\frac{53}{%
		18(s+1)}+\frac{\pi ^{2}}{6(s+1)}-\frac{1}{(s+2)^{3}}\right.  \\
	&&\left. +\frac{5}{6(s+2)^{2}}-\frac{187}{18(s+2)}+\frac{\pi ^{2}}{6(s+2)}-%
	\frac{67}{9}\left( \psi (s+1)+\gamma _{E}\right) +\frac{1}{3}\pi
	^{2}\right.
	\\
	&&\left. \left( \psi (s+1)+\gamma _{E}\right) +2\left( \frac{67}{18}-\frac{%
		\pi ^{2}}{6}\right) \left( \psi (s+1)+\gamma _{E}\right)
	-\frac{11}{3}\psi ^{\prime }(s+1)-\psi ^{\prime \prime
	}(s+1)\right)
\end{eqnarray*}

\begin{eqnarray*}
	&&\Phi _{nsq\overline{q}}^{NLO} =\\
	&&C_{F}\left(
	-\frac{C_{A}}{2}+C_{F}\right)
	\left( \frac{2}{(s+1)^{3}}-\frac{2}{(s+1)^{2}}+\frac{4}{s+1}-\frac{\pi ^{2}}{%
		3(s+1)}-\frac{1.9968}{(s+2)^{3}}-\frac{2}{(s+2)^{2}}\right.  \\
	&&\left. +\frac{3.3246}{s+2}+\frac{3.9404}{(s+3)^{3}}-\frac{7.1312}{s+3}-%
	\frac{3.602}{(s+4)^{3}}+\frac{5.8861}{s+4}+\frac{2.6484}{(s+5)^{3}}+\frac{%
		3.9432}{s+5}-\frac{1.2696}{(s+6)^{3}}\right.  \\
	&&\left. -\frac{14.24}{s+6}+\frac{0.2796}{(s+7)^{3}}+\frac{20.43}{s+7}-\frac{%
		19.77}{s+8}+\frac{13.05}{s+9}+\frac{6.286}{s+10}+\frac{1.997}{s+11}-\frac{%
		0.3076}{s+12}\right.  \\
	&&\left. -2\left( \frac{4}{(s+1)^{3}}-\frac{\ln (4)}{(s+1)^{2}}-\frac{\psi (%
		\frac{s}{2}+1)}{(s+1)^{2}}+\frac{\psi
		(\frac{s+1}{2})}{(s+1)^{2}}+\frac{\psi
		^{\prime }(\frac{s}{2}+1)}{2s+2}-\frac{\psi ^{\prime }(\frac{s+1}{2})}{2(s+1)%
	}\right) \right.  \\
	&&\left. -\frac{0.9984}{(s+2)^{3}}\left( \frac{16}{(s+1)^{2}}+\frac{12s}{%
		(s+1)^{2}}+(s+2)\ln (16)-2(s+2)\psi (\frac{s}{2}+1)+2(s+1)\psi (\frac{s+1}{2}%
	)\right. \right.  \\
	&&\left. \left. +(s+2)^{2}\psi ^{\prime
	}(\frac{s}{2}+1)-(s+2)^{2}\psi
	^{\prime }(\frac{s+1}{2})\right) -\frac{1.9702}{(s+3)^{3}}\left( \frac{164}{%
		(s+1)^{2}(s+2)^{2}}+\right. \right.  \\
	&&\left. \left. \frac{284s}{(s+1)^{2}(s+2)^{2}}+\frac{188s^{2}}{%
		(s+1)^{2}(s+2)^{2}}+\frac{60s^{3}}{(s+1)^{2}(s+2)^{2}}+\frac{8s^{4}}{%
		(s+1)^{2}(s+2)^{2}}-\right. \right.  \\
	&&\left. \left. 4(s+3)\ln (2)-2(s+3)\psi (\frac{s}{2}+1)+2(s+3)\psi (\frac{%
		s+1}{2})+(s+3)^{2}\psi ^{\prime }(\frac{s}{2}+1)-\right. \right.  \\
	&&\left. \left. (s+3)^{2}\psi ^{\prime }(\frac{s+1}{2})\right) -\frac{1.801}{%
		(s+4)^{3}}\left( \frac{2176}{(s+1)^{2}(s+2)^{2}(s+3)^{2}}+\frac{4392s}{%
		(s+1)^{2}(s+2)^{2}(s+3)^{2}}\right. \right.  \\
	&&\left. \left. +\frac{3504s^{2}}{(s+1)^{2}(s+2)^{2}(s+3)^{2}}+\frac{%
		1408s^{3}}{(s+1)^{2}(s+2)^{2}(s+3)^{2}}+\frac{288s^{4}}{%
		(s+1)^{2}(s+2)^{2}(s+3)^{2}}\right. \right.  \\
	&&\left. \left.
	+\frac{24s^{5}}{(s+1)^{2}(s+2)^{2}(s+3)^{2}}+4(s+4)\ln
	(2)-2(s+4)\psi (\frac{s}{2}+1)+2(s+4)\psi (\frac{s+1}{2})+\right. \right.  \\
	&&\left. \left. (s+4)^{2}\psi ^{\prime
	}(\frac{s}{2}+1)-(s+4)^{2}\psi
	^{\prime }(\frac{s+1}{2})\right) -\frac{1.3242}{(s+5)^{3}}\left( \frac{57328%
	}{(s+1)^{2}(s+2)^{2}(s+3)^{2}(s+4)^{2}}+\right. \right.  \\
	&&\left. \left. \frac{146144s}{(s+1)^{2}(s+2)^{2}(s+3)^{2}(s+4)^{2}}+\frac{%
		162160s^{2}}{(s+1)^{2}(s+2)^{2}(s+3)^{2}(s+4)^{2}}+\right. \right.  \\
	&&\left. \left. \frac{103728s^{3}}{(s+1)^{2}(s+2)^{2}(s+3)^{2}(s+4)^{2}}+%
	\frac{42144s^{4}}{(s+1)^{2}(s+2)^{2}(s+3)^{2}(s+4)^{2}}+\right. \right.  \\
	&&\left. \left. \frac{11160s^{5}}{(s+1)^{2}(s+2)^{2}(s+3)^{2}(s+4)^{2}}+%
	\frac{1880s^{6}}{(s+1)^{2}(s+2)^{2}(s+3)^{2}(s+4)^{2}}+\right. \right.  \\
	&&\left. \left. \frac{184s^{7}}{(s+1)^{2}(s+2)^{2}(s+3)^{2}(s+4)^{2}}+\frac{%
		8s^{8}}{(s+1)^{2}(s+2)^{2}(s+3)^{2}(s+4)^{2}}-4(s+5)\ln (2)\right.
	\right.
	\\
	&&\left. \left. -2(5+s)\psi (\frac{s}{2}+1)+2(5+s)\psi (\frac{s+1}{2}%
	)+(s+5)^{2}\psi ^{\prime }(\frac{s}{2}+1)-(s+5)^{2}\psi ^{\prime }(\frac{s+1%
	}{2})\right) -\right.  \\
	&&\left. \frac{0.6348}{(s+6)^{2}}\left( \ln (16)-2\psi
	(\frac{s}{2}+4)+2\psi
	(\frac{s+7}{2})+(s+6)\psi ^{\prime }(\frac{s}{2}+4)-(s+6)\psi ^{\prime }(%
	\frac{s+7}{2})\right) +\right.  \\
	&&\left. \frac{0.1398}{(s+7)^{2}}\left( \ln (16)+2\psi
	(\frac{s}{2}+4)-2\psi
	(\frac{s+9}{2})-(s+7)\psi ^{\prime }(\frac{s}{2}+4)+(s+7)\psi ^{\prime }(%
	\frac{s+9}{2})\right) \right)
\end{eqnarray*}

\begin{eqnarray*}
	&&\Phi _{q}^{NLO} =\\
	&&C_{F}T_{f}\left( -\frac{40}{9s}+\frac{4}{(s+1)^{3}}+\frac{%
		28}{3(s+1)^{2}}-\frac{146}{9(s+1)}+\frac{4}{(s+2)^{3}}+\frac{52}{3(s+2)^{2}}+%
	\frac{94}{9(s+2)}+\right.  \\
	&&\left. \frac{16}{3(s+3)^{2}}+\frac{112}{9(s+3)}+\frac{4}{3}\psi
	^{\prime
	}(s+1)\right) +C_{F}^2\left( \frac{7}{(s+1)^{3}}+\frac{3}{(s+1)^{2}}-\frac{1}{%
	s+1}-\frac{\pi ^{2}}{3(s+1)}+\right.  \\
&&\left. \frac{3.0032}{(s+2)^{3}}+\frac{1}{(s+2)^{2}}+\frac{8.3246}{s+2}+%
\frac{3.9404}{(s+3)^{3}}-\frac{7.1312}{s+3}-\frac{3.602}{(s+4)^{3}}+\frac{%
	5.886}{s+4}+\frac{2.6484}{(s+5)^{3}}\right.  \\
&&\left. +\frac{3.9432}{s+5}-\frac{1.2696}{(s+6)^{3}}-\frac{14.2478}{s+6}+%
\frac{0.2796}{(s+7)^{3}}+\frac{20.4376}{s+7}-\frac{19.7727}{s+8}+\frac{13.056%
}{s+9}-\frac{6.2862}{s+10}\right.  \\
&&\left. +\frac{1.9971}{s+11}-\frac{0.3075}{s+12}-\frac{8}{(s+1)^{3}}+\frac{%
	2\ln (4)}{(s+1)^{2}}+\frac{2\psi (\frac{s}{2}+1)}{(s+1)^{2}}-\frac{2\psi (%
	\frac{s+1}{2})}{(s+1)^{2}}-\frac{\psi ^{\prime }(\frac{s}{2}+1)}{s+1}%
+\right.  \\
&&\left. \frac{\psi ^{\prime }(\frac{s+1}{2})}{(s+1)^{2}}-\frac{0.9984}{%
	(s+2)^{3}}\left(
\frac{16}{(s+1)^{2}}+\frac{12s}{(s+1)^{2}}+(s+2)\ln
(16)-2(s+2)\psi (\frac{s}{2}+1)+\right. \right.  \\
&&\left. \left. 2(s+2)\psi (\frac{s+1}{2})+(s+2)^{2}\psi ^{\prime }(\frac{s}{%
	2}+1)-(s+2)^{2}\psi ^{\prime }(\frac{s+1}{2})\right)- \frac{1.9702}{(s+3)^{3}}%
\left( \frac{164}{(s+1)^{2}(s+2)^{2}}\right. \right.  \\
&&\left. \left. +\frac{284s}{(s+1)^{2}(s+2)^{2}}+\frac{188s^{2}}{%
	(s+1)^{2}(s+2)^{2}}+\frac{60s^{3}}{(s+1)^{2}(s+2)^{2}}+\frac{8s^{4}}{%
	(s+1)^{2}(s+2)^{2}}-\right. \right.  \\
&&\left. \left. 4(s+3)\ln (2)-2(s+3)\psi (\frac{s}{2}+1)+2(s+3)\psi (\frac{%
	s+1}{2})+(s+3)^{2}\psi ^{\prime }(\frac{s}{2}+1)-(s+3)^{2}\psi ^{\prime }(%
\frac{s+1}{2})\right) \right.  \\
&&\left. -\frac{1.801}{(s+4)^{3}}\left( \frac{2176}{%
	(s+1)^{2}(s+2)^{2}(s+3)^{2}}+\frac{4392s}{(s+1)^{2}(s+2)^{2}(s+3)^{2}}+\frac{%
	3504s^{2}}{(s+1)^{2}(s+2)^{2}(s+3)^{2}}+\right. \right.  \\
&&\left. \left. \frac{1408s^{3}}{(s+1)^{2}(s+2)^{2}(s+3)^{2}}+\frac{288s^{4}%
}{(s+1)^{2}(s+2)^{2}(s+3)^{2}}+\frac{24s^{5}}{(s+1)^{2}(s+2)^{2}(s+3)^{2}}%
+\right. \right.  \\
&&\left. \left. 4(s+4)\ln (2)-2(s+4)\psi (\frac{s}{2}+1)+2(s+4)\psi (\frac{%
	s+1}{2})+(s+4)^{2}\psi ^{\prime }(\frac{s}{2}+1)-(s+4)^{2}\psi ^{\prime }(%
\frac{s+1}{2})\right) \right.  \\
&&\left.- \frac{1.3242}{(s+5)^{3}}\left( \frac{57328}{%
	(s+1)^{2}(s+2)^{2}(s+3)^{2}(s+4)^{2}}+\frac{146144s}{%
	(s+1)^{2}(s+2)^{2}(s+3)^{2}(s+4)^{2}}+\right. \right.  \\
&&\left. \left. \frac{162160s^{2}}{(s+1)^{2}(s+2)^{2}(s+3)^{2}(s+4)^{2}}+%
\frac{103728s^{3}}{(s+1)^{2}(s+2)^{2}(s+3)^{2}(s+4)^{2}}+\right. \right.  \\
&&\left. \left. \frac{42144s^{4}}{(s+1)^{2}(s+2)^{2}(s+3)^{2}(s+4)^{2}}+%
\frac{11160s^{5}}{(s+1)^{2}(s+2)^{2}(s+3)^{2}(s+4)^{2}}+\right. \right.  \\
&&\left. \left. \frac{1880s^{6}}{(s+1)^{2}(s+2)^{2}(s+3)^{2}(s+4)^{2}}+\frac{%
	184s^{7}}{(s+1)^{2}(s+2)^{2}(s+3)^{2}(s+4)^{2}}+\right. \right.  \\
&&\left. \left. \frac{8s^{8}}{(s+1)^{2}(s+2)^{2}(s+3)^{2}(s+4)^{2}}%
-4(s+5)\ln (2)-2(s+5)\psi (\frac{s}{2}+1)+2(s+5)\psi
(\frac{s+1}{2})\right.
\right.  \\
&&\left. \left. +(s+5)^{2}\psi ^{\prime
}(\frac{s}{2}+1)-(s+5)^{2}\psi
^{\prime }(\frac{s+1}{2})\right) -\frac{2}{(s+1)^{2}}(\gamma _{E}+\frac{1}{%
	s+1}+\psi (s+1)-(s+1)\psi ^{\prime }(s+2))\right.  \\
&&\left. -\frac{2}{(s+2)^{2}}(\gamma _{E}+\frac{1}{s+2}+\psi
(s+2)-(s+2)\psi
^{\prime }(s+3))-\frac{0.6348}{(s+6)^{2}}\left( \ln (16)-2\psi (\frac{s}{2}%
+4)+\right. \right.  \\
&&\left. \left. 2\psi (\frac{s+7}{2})+(s+6)\psi ^{\prime }(\frac{s}{2}%
+4)-(s+6)\psi ^{\prime }(\frac{s+7}{2})\right) +\frac{0.1398}{(s+7)^{2}}%
\left( \ln (16)+2\psi (\frac{s}{2}+4)-\right. \right.  \\
&&\left. \left. 2\psi (\frac{s+9}{2})-(s+7)\psi ^{\prime }(\frac{s}{2}%
+4)+(s+7)\psi ^{\prime }(\frac{s+9}{2})\right) +\right.  \\
&&\left. 4\left( (\psi (s+1)+\gamma _{E})\psi ^{\prime }(s+1)-\frac{1}{2}%
\psi ^{\prime \prime }(s+1)\right) -3\psi ^{\prime }(s+1)+4\psi
^{\prime
	\prime }(s+1)\right) +\allowdisplaybreaks[1] \\
	&&C_{A}C_{F}\left( \frac{2}{(s+1)^{3}}+\frac{11}{6(s+1)^{2}}+\frac{17}{%
		18(s+1)}+\frac{\pi ^{2}}{3(s+1)}-\frac{0.0016}{(s+2)^{3}}+\frac{11}{%
		6(s+2)^{2}}-\frac{10.4062}{s+2}\right.  \\
	&&\left. -\frac{1.9702}{(s+3)^{3}}+\frac{3.5656}{s+3}+\frac{1.801}{(s+4)^{3}}%
	-\frac{2.9430}{s+4}-\frac{1.3242}{(s+5)^{3}}-\frac{1.9716}{s+5}+\frac{0.6348%
	}{(s+6)^{3}}+\frac{7.1239}{s+6}\right.  \\
	&&\left. -\frac{0.1398}{(s+7)^{3}}-\frac{10.2188}{s+7}+\frac{9.8863}{s+8}-%
	\frac{6.5284}{s+9}+\frac{3.1431}{s+10}-\frac{0.9985}{s+11}+\frac{0.1537}{s+12%
	}-\right.  \\
	&&\left. \frac{67(\psi (s+1)+\gamma _{E})}{9}+\frac{1}{3}\pi
	^{2}(\psi (s+1)+\gamma _{E})+2\left( \frac{67}{18}-\frac{\pi
		^{2}}{6}\right) (\psi
	(s+1)+\gamma _{E})-\frac{\ln (4)}{(s+1)^{2}}-\right.  \\
	&&\left. \frac{\psi (\frac{s}{2}+1)}{(s+1)^{2}}+\frac{\psi (\frac{s+1}{2})}{%
		(s+1)^{2}}+\frac{\psi ^{\prime }(\frac{s}{2}+1)}{2s+2}-\frac{\psi ^{\prime }(%
		\frac{s+1}{2})}{2s+2}+\frac{0.4992}{(s+2)^{3}}\left( \frac{16}{(s+1)^{2}}+%
	\frac{12s}{(s+1)^{2}}+(s+2)\ln (16)\right. \right.  \\
	&&\left. \left. -2(s+2)\psi (\frac{s}{2}+1)+2(s+2)\psi (\frac{s+1}{2}%
	)+(s+2)^{2}\psi ^{\prime }(\frac{s}{2}+1)-(s+2)^{2}\psi ^{\prime }(\frac{s+1%
	}{2})\right) +\right.  \\
	&&\left. \frac{0.9851}{(s+3)^{3}}\left( \frac{164}{(s+1)^{2}(s+2)^{2}}+\frac{%
		284s}{(s+1)^{2}(s+2)^{2}}+\frac{188s^{2}}{(s+1)^{2}(s+2)^{2}}+\frac{60s^{3}}{%
		(s+1)^{2}(s+2)^{2}}+\right. \right.  \\
	&&\left. \left. \frac{8s^{4}}{(s+1)^{2}(s+2)^{2}}-4(s+3)\ln (2)-2(s+3)\psi (%
	\frac{s}{2}+1)+2(s+3)\psi (\frac{s+1}{2})+\right. \right.  \\
	&&\left. \left. (s+3)^{2}\psi ^{\prime
	}(\frac{s}{2}+1)-(s+3)^{2}\psi
	^{\prime }(\frac{s+1}{2})\right) +\frac{0.9005}{(s+4)^{3}}\right.  \\
	&&\left. \left( \frac{2176}{(s+1)^{2}(s+2)^{2}(s+3)^{2}}+\frac{4392s}{%
		(s+1)^{2}(s+2)^{2}(s+3)^{2}}+\frac{3504s^{2}}{(s+1)^{2}(s+2)^{2}(s+3)^{2}}%
	+\right. \right.  \\
	&&\left. \left. \frac{1408s^{3}}{(s+1)^{2}(s+2)^{2}(s+3)^{2}}+\frac{288s^{4}%
	}{(s+1)^{2}(s+2)^{2}(s+3)^{2}}+\frac{24s^{5}}{(s+1)^{2}(s+2)^{2}(s+3)^{2}}%
	+\right. \right.  \\
	&&\left. \left. 4(s+4)\ln (2)-2(s+4)\psi (\frac{s}{2}+1)+2(s+4)\psi (\frac{%
		s+1}{2})+(s+4)^{2}\psi ^{\prime }(\frac{s}{2}+1)-(s+4)^{2}\psi ^{\prime }(%
	\frac{s+1}{2})\right) \right.  \\
	&&\left. +\frac{0.6621}{(s+5)^{3}}\left( \frac{57328}{%
		(s+1)^{2}(s+2)^{2}(s+3)^{2}(s+4)^{2}}+\frac{146144s}{%
		(s+1)^{2}(s+2)^{2}(s+3)^{2}(s+4)^{2}}+\right. \right.  \\
	&&\left. \left. \frac{162160s^{2}}{(s+1)^{2}(s+2)^{2}(s+3)^{2}(s+4)^{2}}+%
	\frac{103728s^{3}}{(s+1)^{2}(s+2)^{2}(s+3)^{2}(s+4)^{2}}+\right. \right.  \\
	&&\left. \left. \frac{42144s^{4}}{(s+1)^{2}(s+2)^{2}(s+3)^{2}(s+4)^{2}}+%
	\frac{11160s^{5}}{(s+1)^{2}(s+2)^{2}(s+3)^{2}(s+4)^{2}}+\right. \right.  \\
	&&\left. \left. \frac{1880s^{6}}{(s+1)^{2}(s+2)^{2}(s+3)^{2}(s+4)^{2}}+\frac{%
		184s^{7}}{(s+1)^{2}(s+2)^{2}(s+3)^{2}(s+4)^{2}}+\right. \right.  \\
	&&\left. \left. \frac{8s^{8}}{(s+1)^{2}(s+2)^{2}(s+3)^{2}(s+4)^{2}}%
	-4(s+5)\ln (2)-2(s+5)\psi (\frac{s}{2}+1)+2(s+5)\psi
	(\frac{s+1}{2})+\right.
	\right.  \\
	&&\left. \left. (s+5)^{2}\psi ^{\prime
	}(\frac{s}{2}+1)-(s+5)^{2}\psi ^{\prime }(\frac{s+1}{2})\right)
	+\frac{0.3174}{(s+6)^{2}}\left( \ln
	(16)-2\psi (\frac{s}{2}+4)+2\psi (\frac{s+7}{2})+\right. \right.  \\
	&&\left. \left. (s+6)\psi ^{\prime }(\frac{s}{2}+4)-(s+6)\psi ^{\prime }(%
	\frac{s+7}{2})\right) -\frac{0.0699}{(s+7)^{2}}\left( \ln (16)+2\psi (\frac{s%
	}{2}+4)-2\psi (\frac{s+9}{2})-\right. \right.  \\
	&&\left. \left. (s+7)\psi ^{\prime }(\frac{s}{2}+4)+(s+7)\psi ^{\prime }(%
	\frac{s+9}{2})\right) -\frac{11}{3}\psi ^{\prime }(s+1)-\psi
	^{\prime \prime }(s+1)\right)
\end{eqnarray*}

\begin{eqnarray*}
	&&\Theta _{q}^{NLO} =\\
	&&T_{f}^2\left( \frac{8}{3(s+1)^{2}}-\frac{40}{9(s+1)}-%
	\frac{16}{3(s+2)^{2}}+\frac{32}{9(s+2)}+\frac{16}{3(s+3)^{ý2ý}}-\frac{32}{%
		9(s+3)}+\right.  \\
	&&\left. \frac{8(\psi (s+2)+\gamma _{E})}{3(s+1)}-\frac{16(\psi
		(s+3)+\gamma
		_{E})}{3(s+2)}+\frac{16(\psi (s+4)+\gamma _{E})}{3(s+3)}\right) + \\
	&&C_{A}T_{f}\left( -\frac{40}{9s}+\frac{4}{(s+1)^{3}}+\frac{8}{3(s+1)^{2}}+%
	\frac{26}{9(s+1)}+\frac{24}{(s+2)^{3}}+\frac{68}{3(s+2)^{2}}-\frac{33.231}{%
		s+2}\right.  \\
	&&\left. -\frac{4\pi ^{2}}{3(s+2)}+\frac{8}{3(s+3)^{2}}+\frac{96.875}{s+3}-%
	\frac{67.644}{s+4}+\frac{83.04}{s+5}-\frac{82.976}{s+6}+\frac{56.16}{s+7}-%
	\frac{22}{s+8}\right.  \\
	&&\left. -\frac{22(\psi (s+2)+\gamma _{E})}{3(s+1)}+\frac{20(\psi
		(s+3)+\gamma _{E})}{3(s+2)}-\frac{20(\psi (s+4)+\gamma
		_{E})}{3(s+3)}\right.
	\\
	&&\left. -2\left( \frac{4}{(s+1)^{3}}-\frac{\ln (4)}{(s+1)^{2}}-\frac{\psi (%
		\frac{s}{2}+1)}{(s+1)^{2}}+\frac{\psi
		(\frac{s+1}{2})}{(s+1)^{2}}+\frac{\psi
		^{\prime }(\frac{s}{2}+1)}{2s+2}-\frac{\psi ^{\prime }(\frac{s+1}{2})}{2(s+1)%
	}\right) +\right.  \\
	&&\left. \frac{2}{(s+2)^{3}}\left( \frac{16}{(s+1)^{2}}+\frac{12s}{(s+1)^{2}}%
	+(s+2)\ln (16)-2(s+2)\psi (\frac{s}{2}+1)+2(s+2)\psi
	(\frac{s+1}{2})\right.
	\right.  \\
	&&\left. \left. +(s+2)^{2}\psi ^{\prime
	}(\frac{s}{2}+1)-(s+2)^{2}\psi
	^{\prime }(\frac{s+1}{2})\right) -\frac{2}{(s+3)^{3}}\left( \frac{164}{%
		(s+1)^{2}(s+2)^{2}}+\frac{284s}{(s+1)^{2}(s+2)^{2}}\right. \right.  \\
	&&\left. \left. +\frac{188s^{2}}{(s+1)^{2}(s+2)^{2}}+\frac{60s^{3}}{%
		(s+1)^{2}(s+2)^{2}}+\frac{8s^{4}}{(s+1)^{2}(s+2)^{2}}-4(s+3)\ln
	(2)-2(s+3)\psi (\frac{s}{2}+1)ý+ý\right. \right.  \\
	&&\left. \left. 2(s+3)\psi (\frac{s+1}{2})+(s+ý3ý)^{2}\psi ^{\prime }(\frac{s}{%
		2}+1)-(s+3)^{2}\psi ^{\prime }(\frac{s+1}{2})\right) +\frac{2(\pi
		^{2}+6(\psi (s+2)+\gamma _{E})^{2}-6\psi ^{\prime }(s+2))}{6s+6}\right.  \\
	&&\left. -\frac{8}{(s+1)^{2}}\left( \gamma _{E}+\frac{1}{s+1}+\psi
	(s+1)-(s+1)\psi ^{\prime }(s+2)\right) -\frac{4(\pi ^{2}+6(\psi
		(s+3)+\gamma
		_{E})^{2}-6\psi ^{\prime }(s+3))}{6s+12}\right.  \\
	&&\left. +\frac{16}{(s+2)^{2}}\left( \gamma
	_{E}+\frac{1}{s+2}+\psi (s+2)-(s+2)\psi ^{\prime }(s+3)\right)
	+\frac{4(\pi ^{2}+6(\psi (s+4)+\gamma
		_{E})^{2}-6\psi ^{\prime }(s+4))}{6s+18}\right.  \\
	&&\left. -\frac{16}{(s+3)^{2}}\left( \gamma
	_{E}+\frac{1}{s+3}+\psi
	(s+3)-(s+3)\psi ^{\prime }(s+4)\right) \right) + \\
	&&C_{F}T_{f}\left( -\frac{2}{(s+1)^{3}}+\frac{7}{(s+1)^{2}}-\frac{12}{s+1}-%
	\frac{2\pi ^{2}}{3(s+1)}+\frac{4}{(s+2)^{3}}-\frac{8}{(s+2)^{2}}+\frac{39.16%
	}{s+2}+\frac{4\pi ^{2}}{3(s+2)}-\frac{8}{(s+ý3ý)^{3}}\right.  \\
	&&\left. -\frac{65.856}{s+3}-\frac{4\pi ^{2}}{3(s+3)}+\frac{77.872}{s+4}-%
	\frac{81.216}{s+5}+\frac{80.128}{s+6}-\frac{51.968}{s+7}+\frac{17.6}{s+8}+%
	\frac{2(\psi (s+1)+\gamma _{E})}{s+1}+\right.  \\
	&&\left. \frac{4(\psi (s+2)+\gamma _{E})}{s+1}-\frac{4(\psi
		(s+2)+\gamma _{E})}{s+2}+\frac{4(\psi (s+3)+\gamma
		_{E})}{s+3}-\frac{2(\pi ^{2}+6(\psi
		(s+2)+\gamma _{E})^{2}-6\psi ^{\prime }(s+2))}{6s+6}\right.  \\
	&&\left. +\frac{12}{(s+1)^{2}}\left( \gamma
	_{E}+\frac{1}{s+1}+\psi (s+1)-(s+1)\psi ^{\prime }(s+2)\right)
	+\frac{4(\pi ^{2}+6(\psi (s+3)+\gamma
		_{E})^{2}-6\psi ^{\prime }(s+3))}{6s+12}\right.  \\
	&&\left. -\frac{24}{(s+2)^{2}}\left( \gamma
	_{E}+\frac{1}{s+2}+\psi (s+2)-(s+2)\psi ^{\prime }(s+3)\right)
	-\frac{4(\pi ^{2}+6(\psi (s+4)+\gamma
		_{E})^{2}-6\psi ^{\prime }(s+4))}{6s+18}+\right.  \\
	&&\left. \frac{24}{(s+3)^{2}}\left( \gamma _{E}+\frac{1}{s+3}+\psi
	(s+3)-(s+3)\psi ^{\prime }(s+4)\right) \right)
\end{eqnarray*}

\begin{eqnarray*}
	&&\Theta _{g}^{NLO} =\\
	&&C_{F}^{2}\left( \frac{2}{(s+1)^{3}}+\frac{8}{(s+1)^{2}}-%
	\frac{16.66}{s+1}-\frac{1}{(s+2)^{3}}-\frac{1}{2(s+2)^{2}}+\frac{34.196}{s+2}%
	-\frac{40.096}{s+3}+\right.  \\
	&&\left. \frac{42.432}{s+4}-\frac{35.224}{s+5}+\frac{17.392}{s+6}-\frac{4.4}{%
		s+7}-\frac{2(\psi (s+3)+\gamma _{E})}{s+2}+\right.  \\
	&&\left. \frac{1}{3s}\left( \pi ^{2}+6(\psi (s+1)+\gamma
	_{E})^{2}-6\psi ^{\prime }(s+1)\right) -\frac{8}{s^{3}}(1+s\gamma
	_{E}+s(\psi (s)-s\psi
	^{\prime }(s+1)))-\right.  \\
	&&\left. \frac{2}{6s+6}\left( \pi ^{2}+6\left( \psi (s+2)+\gamma
	_{E}\right)
	^{2}-6\psi ^{\prime }(s+2)\right) +\frac{8}{(s+1)^{2}}\left( \gamma _{E}+%
	\frac{1}{s+1}+\psi (s+1)-\right. \right.  \\
	&&\left. \left. (s+1)\psi ^{\prime }(s+2)\right)
	+\frac{1}{6s+12}\left( \pi
	^{2}+6(\psi (s+3)+\gamma _{E})^{2}-6\psi ^{\prime }(s+3)\right) -\right.  \\
	&&\left. \frac{4}{(s+2)^{2}}\left( \gamma _{E}+\frac{1}{s+2}+\psi
	(s+2)-(s+2)\psi ^{\prime }(s+3)\right) \right) + \\
	&&C_{A}C_{F}\left(
	-\frac{4}{s^{3}}+\frac{6}{s^{2}}+\frac{17}{9s}-\frac{2\pi
		^{2}}{3s}-\frac{8}{(s+1)^{2}}+\frac{25.2}{s+1}-\frac{4}{(s+2)^{3}}-\frac{9}{%
		(s+2)^{2}}-\frac{23.27}{s+2}-\right.  \\
	&&\left. \frac{\pi ^{2}}{3(s+2)}-\frac{8}{3(s+3)^{2}}+\frac{35.99}{s+3}-%
	\frac{41.046}{s+4}+\frac{35.01}{s+5}-\frac{17.444}{s+6}+\frac{3.3}{s+7}+%
	\frac{2(\psi (s+3)+\gamma _{E})}{s+2}\right.  \\
	&&\left. +\frac{1}{s^{2}}\left( \ln (16)-2\psi (\frac{s}{2}+1)+2\psi (\frac{%
		s+1}{2})+s\psi ^{\prime }(\frac{s}{2}+1)-s\psi ^{\prime }(\frac{s+1}{2}%
	)\right) -\right.  \\
	&&\left. 2\left( \frac{4}{(s+1)^{3}}-\frac{\ln (4)}{(s+1)^{2}}-\frac{\psi (%
		\frac{s}{2}+1)}{(s+1)^{2}}+\frac{\psi
		(\frac{s+1}{2})}{(s+1)^{2}}+\frac{\psi
		^{\prime }(\frac{s}{2}+1)}{2s+2}-\frac{\psi ^{\prime }(\frac{s+1}{2})}{2s+2}%
	\right) +\right.  \\
	&&\left. \frac{1}{2(s+2)^{3}}\left( \frac{16}{(s+1)^{2}}+\frac{12s}{(s+1)^{2}%
	}+(s+2)\ln (16)-2(s+2)\psi (\frac{s}{2}+1)+2(s+2)\psi (\frac{s+1}{2}%
	)+\right. \right.  \\
	&&\left. \left. (s+2)^{2}\psi ^{\prime
	}(\frac{s}{2}+1)-(s+2)^{2}\psi
	^{\prime }(\frac{s+1}{2})\right) -\frac{1}{3s}\right.  \\
	&&\left. \left( \pi ^{2}+6\left( \psi (s+1)+\gamma _{E}\right)
	^{2}-6\psi ^{\prime }(s+1)\right) +\frac{12}{s^{3}}(1+s\gamma
	_{E}+s(\psi (s)-s\psi
	^{\prime }(s+1)))+\right.  \\
	&&\left. \frac{2}{6s+6}\left( \pi ^{2}+6\left( \psi (s+2)+\gamma
	_{E}\right)
	^{2}-6\psi ^{\prime }(s+2)\right) -\right.  \\
	&&\left. \frac{12}{(s+1)^{2}}\left( \gamma _{E}+\frac{1}{s+1}+\psi
	(s+1)-(s+1)\psi ^{\prime }(s+2)\right) -\right.  \\
	&&\left. \frac{1}{6s+12}\left( \pi ^{2}+6\left( \psi (s+3)+\gamma
	_{E}\right) ^{2}-6\psi ^{\prime }(s+3)\right) +\right.  \\
	&&\left. \frac{6}{(s+2)^{2}}\left( \gamma _{E}+\frac{1}{s+2}+\psi
	(s+2)-(s+2)\psi ^{\prime }(s+3)\right) \right)
\end{eqnarray*}

\begin{eqnarray*}
	&&\Phi _{g}^{NLO} =\\
	&&C_{F}T_{f}\left( -\frac{16}{3s^{2}}+\frac{92}{9s}+\frac{4%
	}{(s+1)^{3}}-\frac{10}{(s+1)^{2}}-\frac{4}{s+1}+\frac{4}{(s+2)^{3}}-\frac{14%
}{(s+2)^{2}}+\frac{12}{s+2}-\frac{16}{3(s+3)^{2}}-\frac{164}{9(s+3)}\right)  \\
&&+C_{A}T_{f}\left( \frac{8}{3s^{2}}-\frac{46}{9s}-\frac{4}{(s+1)^{2}}+\frac{%
	58}{9(s+1)}+\frac{4}{(s+2)^{2}}-\frac{38}{9(s+2)}-\frac{8}{3(s+3)^{2}}+\frac{%
	46}{9(s+3)}\right.  \\
&&\left. +\frac{8}{3}\psi ^{\prime }(s+1)\right) +C_{A}^{2}\left( -\frac{8}{%
	s^{3}}+\frac{22}{3s^{2}}+\frac{2}{(s+1)^{3}}+\frac{11}{(s+1)^{2}}+\frac{%
	4.4407}{s+1}-\frac{17.9984}{(s+2)^{3}}+\frac{1}{(s+2)^{2}}-\right.  \\
&&\left. \frac{6.9024}{s+2}-\frac{\pi ^{2}}{3(s+2)}+\frac{5.9702}{(s+3)^{3}}+%
\frac{22}{3(s+3)^{2}}-\frac{6.7917}{s+3}+\frac{\pi ^{2}}{3(s+3)}-\frac{1.801%
}{(s+4)^{3}}-\frac{3.5389}{s+4}+\right.  \\
&&\left. \frac{1.3242}{(s+5)^{3}}+\frac{1.2736}{s+5}-\frac{0.6348}{(s+6)^{3}}%
-\frac{5.6479}{s+6}+\frac{0.1398}{(s+7)^{3}}+\frac{9.2228}{s+7}-\frac{7.6863%
}{s+8}+\frac{6.5284}{s+9}-\frac{3.1431}{s+10}\right.  \\
&&\left. +\frac{0.9985}{s+11}-\frac{0.1537}{s+12}-\frac{67(\psi
	(s+1)+\gamma _{E})}{9}+\frac{1}{3}\pi^2(\psi (s+1)+\gamma
_{E})-\frac{1}{s^{2}}\left( \ln
(16)-2\psi (\frac{s}{2}+1)+\right. \right.  \\
&&\left. \left. 2\psi (\frac{s+1}{2})+s\psi ^{\prime
}(\frac{s}{2}+1)-s\psi
^{\prime }(\frac{s+1}{2})\right) +2\left( \frac{4}{(s+1)^{3}}-\frac{\ln (4)}{%
	(s+1)^{2}}-\frac{\psi (\frac{s}{2}+1)}{(s+1)^{2}}+\frac{\psi (\frac{s+1}{2})%
}{(s+1)^{2}}+\right. \right.  \\
&&\left. \left. \frac{\psi ^{\prime
	}(\frac{s}{2}+1)}{2s+2}-\frac{\psi ^{\prime
}(\frac{s+1}{2})}{2s+2}\right) -\frac{1.9992}{(s+2)^{3}}\left(
\frac{16}{(s+1)^{2}}+\frac{12s}{(s+1)^{2}}+(s+2)\ln (16)-2(s+2)\psi (\frac{s%
}{2}+1)\right. \right.  \\
&&\left. \left. +2(s+2)\psi (\frac{s+1}{2})+(s+2)^{2}\psi ^{\prime }(\frac{s%
}{2}+1)-(s+2)^{2}\psi ^{\prime }(\frac{s+1}{2})\right) +\frac{0.0149}{%
(s+1)^{3}}\left( \frac{164}{(s+1)^{2}(s+2)^{2}}\right. \right.  \\
&&\left. \left. +\frac{284s}{(s+1)^{2}(s+2)^{2}}+\frac{188s^{2}}{%
	(s+1)^{2}(s+2)^{2}}+\frac{60s^{3}}{(s+1)^{2}(s+2)^{2}}+\frac{8s^{4}}{%
	(s+1)^{2}(s+2)^{2}}-4(s+3)\ln (2)-\right. \right.  \\
&&\left. \left. 2(s+3)\psi (\frac{s}{2}+1)+2(s+3)\psi (\frac{s+1}{2}%
)+(s+3)^{2}\psi ^{\prime }(\frac{s}{2}+1)-(s+3)^{2}\psi ^{\prime }(\frac{s+1%
}{2})\right) -\frac{0.9005}{(s+4)^{3}}\right.  \\
&&\left. \left( \frac{2176}{(s+1)^{2}(s+2)^{2}(s+3)^{2}}+\frac{4392s}{%
	(s+1)^{2}(s+2)^{2}(s+3)^{2}}+\frac{3504s^{2}}{(s+1)^{2}(s+2)^{2}(s+3)^{2}}%
\right. \right.  \\
&&\left. \left. +\frac{1408s^{3}}{(s+1)^{2}(s+2)^{2}(s+3)^{2}}+\frac{288s^{4}%
}{(s+1)^{2}(s+2)^{2}(s+3)^{2}}+\frac{24s^{5}}{(s+1)^{2}(s+2)^{2}(s+3)^{2}}%
+\right. \right.  \\
&&\left. \left. 4(s+4)\ln (2)-2(s+4)\psi (\frac{s}{2}+1)+2(s+4)\psi (\frac{%
	s+1}{2})+(s+4)^{2}\psi ^{\prime }(\frac{s}{2}+1)-(s+4)^{2}\psi ^{\prime }(%
\frac{s+1}{2})\right) \right.  \\
&&\left. -\frac{0.6621}{(s+5)^{3}}\left( \frac{57328}{%
	(s+1)^{2}(s+2)^{2}(s+3)^{2}(s+4)^{2}}+\frac{146144s}{%
	(s+1)^{2}(s+2)^{2}(s+3)^{2}(s+4)^{2}}\right. \right.  \\
&&\left. +\frac{162160s^{2}}{(s+1)^{2}(s+2)^{2}(s+3)^{2}(s+4)^{2}}\left. +%
\frac{103728s^{3}}{(s+1)^{2}(s+2)^{2}(s+3)^{2}(s+4)^{2}}\right. \right.  \\
&&\left. \left. +\frac{42144s^{4}}{(s+1)^{2}(s+2)^{2}(s+3)^{2}(s+4)^{2}}\left. +%
\frac{11160s^{5}}{(s+1)^{2}(s+2)^{2}(s+3)^{2}9s+4)^{2}}\right.
\right.
\right.  \\
&&\left. \left. +\frac{1880s^{6}}{(s+1)^{2}(s+2)^{2}(s+3)^{2}(s+4)^{2}}+\frac{%
	184s^{7}}{(s+1)^{2}(s+2)^{2}(s+3)^{2}(s+4)^{2}}+\frac{8s^{8}}{%
	(s+1)^{2}(s+2)^{2}(s+3)^{2}(s+4)^{2}}\right. \right.  \\
&&\left. -4(s+5)\ln (2)-2(s+5)\psi (\frac{s}{2}+1)+2(s+5)\psi (\frac{s+1}{2}%
)+(s+5)^{2}\psi ^{\prime }(\frac{s}{2}+1)-(s+5)^{2}\psi ^{\prime }(\frac{s+1%
}{2})\right)\allowdisplaybreaks[1]  \\
&&\left. +\frac{4}{s^{3}}(1+\gamma _{E}s+s(\psi (s)-s\psi ^{\prime }(s+1)))-%
\frac{8}{(s+1)^{2}}(\gamma _{E}+\frac{1}{s+1}+\psi (s+1)-(s+1)\psi
^{\prime
}(s+2))+\right.  \\
&&\left. \frac{4}{(s+2)^{2}}(\gamma _{E}+\frac{1}{s+2}+\psi
(s+2)-(s+2)\psi ^{\prime }(s+3))-\frac{4}{(s+3)^{2}}(\gamma
_{E}+\frac{1}{s+3}+\psi
(s+3)-(s+3)\psi ^{\prime }(s+4))\right)  \\
&&\left. -\frac{0.3174}{(s+6)^{2}}\left( \ln (16)-2\psi
(\frac{s}{2}+4)+2\psi
(\frac{s+7}{2})+(s+6)\psi ^{\prime }(\frac{s}{2}+4)-(s+6)\psi ^{\prime }(%
\frac{s+7}{2})\right) +\right.  \\
&&\left. \frac{0.0699}{(s+7)^{2}}\left( \ln (16)+2\psi
(\frac{s}{2}+4)-2\psi
(\frac{s+9}{2})-(s+7)\psi ^{\prime }(\frac{s}{2}+4)+(s+7)\psi ^{\prime }(%
\frac{s+9}{2})\right) \right.  \\
&&\left. +4\left( (\psi (s+1)+\gamma _{E})\psi ^{\prime }(s+1)-\frac{1}{2}%
\psi ^{\prime \prime }(s+1)\right) -\frac{22}{3}\psi ^{\prime
}(s+1)+3\psi ^{\prime \prime }(s+1)\right)
\end{eqnarray*}


\end{document}